\documentstyle[12pt]{article}
\begin{document}
\begin{center}
{\bf THE MODEL OF THE UNIVERSE WITH TWO SPACES}
\end{center}
\medskip

D.L.Khokhlov
\medskip

Sumy State University, R.-Korsakov St. 2, Sumy 244007 Ukraine

e-mail: nik@demex.sumy.ua

\medskip
\medskip
\medskip
\medskip

\begin{tabular}{p{5.8in}}
{\small
The model of the homogenous and isotropic universe
with two spaces is considered.
The background space is a coordinate system of reference
and defines the behaviour of the universe.
The other space characterizes the gravity
of the matter of the universe.
In the presented model,
the first derivative of the scale factor
of the universe with respect to time
is equal to the velocity of light.
The density of the matter of the universe
changes from the Plankian value at the Planck time
to the modern value at the modern time.
The model under consideration describes the
universe from the Planck time to the modern time and avoids the
problems of the Friedmann model such as the flatness problem and
the horizon problem.
}
\end{tabular}

\medskip
\medskip
\medskip
\centerline {\bf 1. Introduction}
\medskip

As known \cite{Dol,Lin}, the Friedmann model of the universe
has fundamental
difficulties such as the flatness problem and the horizon problem.
These appears to be a consequence of that the space of the
Friedmann universe, on the one hand, is defined by the gravity
of the matter of the universe, and on the other hand,
is a coordinate system of reference. The solution of the problem
is to introduce the background space as a coordinate system of
reference. In this case,
the background space defines the behaviour of the universe,
and the other space characterizes the gravity
of the matter of the universe.

\medskip
\centerline {\bf 2. Theory}
\medskip

Let us consider the model of the homogenous and isotropic
universe with two spaces.
Let us introduce the background space as a coordinate system of
reference. Then the evolution of the universe is described as
a deformation of the background space.
Let us take the homogenous and isotropic background space,
with the spatial interval of its metric is given by
\begin{equation}
d\hat s^2={{a^{2}d\tilde l^2}\over
\displaystyle\left[1+{{k\tilde l^2}\over 4}\right]^2}
\label{eq:b}
\end{equation}

Suppose that the background space is defined by the total mass
of the universe including the mass of the matter and
the energy of gravity
\begin{equation}
\hat G_{ik}=T_{ik}^{tot}=T_{ik}+t_{ik}.\label{eq:c}
\end{equation}
Let us consider the case when the total mass of the universe
is equal to zero
\begin{equation}
T_{ik}^{tot}=T_{ik}+t_{ik}=0.\label{eq:d}
\end{equation}
Then eq. (\ref{eq:c}) take the form
\begin{equation}
\hat G_{ik}=0.\label{eq:e}
\end{equation}
The solution of the equations (\ref{eq:e}) gives
\begin{equation}
{{d^2a}\over{dt^2}}=0\label{eq:f}
\end{equation}
\begin{equation}
{{da}\over{dt}}=c.\label{eq:g}
\end{equation}
Thus
the second derivative of the scale factor of the universe
with respect to time is equal to zero, and the first derivative
of the scale factor is equal to the velocity of light.
It should be noted that
the scale factor of the universe
coincides with the size of the horizon
\begin{equation}
a=ct.\label{eq:g1}
\end{equation}

In the model (\ref{eq:e}), the laboratory coordinate system
is synchronous.
In the laboratory coordinate system, the background space
is described by the flat metric
\begin{equation}
d\hat s^2=c^2dt^2-a^2 d\tilde l^2.\label{eq:h}
\end{equation}
Thus we arrive at the Milne model \cite{Zeld}
in which the size of the universe
being the maximum distance between the particles
coincides with the scale factor of the universe
and coincides with the size of the horizon.
In the universe with one space, the Milne model
is derived from the condition that the density of the matter
tends to zero $\rho\rightarrow 0$. Here the Milne model
describes the background space of
the universe, with the total mass of the universe being equal
to zero $m_{tot}=0$.

Let us determine the relationship between
the lifetime of the universe and the Hubble constant.
Since the Hubble constant is
\begin{equation}
H={1\over a}{da\over dt},\label{eq:j}
\end{equation}
so from (\ref{eq:g}), (\ref{eq:g1}), (\ref{eq:j}) one can obtain
\begin{equation}
t={1\over H}.\label{eq:k}
\end{equation}

Let us estimate the size of the universe at the Planck time
and at present.
Remind that the size of the universe
coincides with the scale factor of the universe.
According to (\ref{eq:g1}),
at the Planck time $t_{Pl}=(\hbar G/c^5)^{1/2}$,
the scale factor of the universe
is equal to the Planck length $l_{Pl}=(\hbar G/c^3)^{1/2}$.
According to (\ref{eq:g1}), (\ref{eq:k}),
for the modern Hubble constant
$H_0 \approx 3 \cdot 10^{-18}{\ \rm c^{-1}}$,
the modern scale factor of the universe is
$a_0 \approx 10^{28}{\ \rm cm}$.

Let us determine the relationship between the mass of the matter
and the scale factor of the universe at $t={\rm const}$.
The total mass of the universe is equal to zero, given
the mass of the matter is equal to the energy of its gravity
\begin{equation}
m={Gm\over{c^2a}}.\label{eq:o}
\end{equation}
Allowing for (\ref{eq:g1}) and (\ref{eq:k}),
from (\ref{eq:o}) it follows that
the mass of the matter changes with time as
\begin{equation}
m={c^2a\over{G}}={c^3t\over{G}}={c^3\over{GH}},\label{eq:p}
\end{equation}
and the density of the matter, as
\begin{equation}
\rho={{3c^2}\over{4\pi G a^2}}={3\over{4\pi G t^2}}=
{{3H^2}\over{4\pi G}}.\label{eq:q}
\end{equation}
According to (\ref{eq:p}), growth of the mass of the matter takes place
during all the evolution of the universe. At the Planck time $t_{Pl}$,
the mass of the matter is equal to the Planck mass
$m_{Pl}=(\hbar c/G)^{1/2}$. At present, the mass of the matter is
$m_0 \approx 1.4 \cdot 10^{56}{\ \rm g}$, and the density of the matter
is $\rho_0 \approx 3.2 \cdot 10^{-29}{\ \rm g\ cm^{-3}}$.
Thus the model of the universe (\ref{eq:d})-(\ref{eq:g})
provides growth of the
mass of the matter from the Plankian value to the modern one.

\medskip
\centerline {\bf 3. Conclusion}
\medskip

We have considered the model of the homogenous and isotropic
universe with two spaces,
with the behaviour of the universe is defined by
the background space. Unlike the Friedmann model, the presented
model gets rid off the flatness and horizon problems.

Remind \cite{Dol,Lin} that
the horizon problem
in the Friedmann universe is that
two particles situated within the horizon at present
were situated beyond the horizon in the past.
In the universe under consideration,
all the particles are situated within the horizon
during all the evolution of the universe,
since the size of the universe
being the maximum distance between the particles
coincides with the size of the horizon.
Hence the presented model avoids the horizon problem.

Remind \cite{Dol,Lin} that
the essence of the flatness problem
in the Friedmann universe
is impossibility to get the
modern density of the matter starting from
the Planck density of the matter at the Planck time.
In the presented theory,
the density of the matter of the universe
changes from the Plankian value at the Planck time
to the modern value at the modern time.
Hence the flatness problem is absent in the presented theory.

In order to resolve the above problems of the Friedmann universe
an inflationary episode is
introduced in the early universe~\cite{Dol,Lin}.
Since the presented model describes the universe
from the Planck time to the modern time and
avoids the above problems of the Friedmann universe,
there is no necessity to introduce
the inflationary model.

\end{document}